\documentclass{JINST}

\usepackage{pifont}

\newcommand{\Ei}{$E_{\rm i}$}
\newcommand{\Ed}{$E_{\rm d}$}
\newcommand{\um}{\Pisymbol{psy}{"6D}m} 
\newcommand{\us}{\Pisymbol{psy}{"6D}s} 
 
\newcommand{\dV}{d$V_{\rm GEM}$}

\title{Property of LCP-GEM in Pure Dimethyl Ether at Low Pressure}

\author{Y. Takeuchi$^{a,b}$, T. Tamagawa$^{a,b}$, T. Kitaguchi$^a$, S. Yamada$^a$, W. Iwakiri$^a$, F. Asami$^{a,b}$, A.~Yoshikawa$^{a,b}$, 
K. Kaneko$^{a,b}$, T. Enoto$^{a,c}$, A. Hayato$^{a}$,  T.Kohmura$^{d}$, and the~GEMS/XACT~team \\
\llap{$^a$}RIKEN Nishina Center, \\
 2-1 Hirosawa, Wako, Saitama, 351-0198, Japan \\
\llap{$^b$} Tokyo Univ. of Science, \\
 1-3 Kagurazaka, Shinjuku-ku, Tokyo, 162-8601, Japan  \\
\llap{$^c$} NASA/GSFC,  \\
8800 Greenbelt Rd, Greenbelt, Maryland 20771, USA \\
\llap{$^d$} Kogakuin Univ., \\
2665-1 Nakano-machi, Hachioji-shi, Tokyo 192-0015, Japan \\
E-mail: \email{takeuhchi@crab.riken.jp}}

\abstract{
We present a systematic investigation of the gain properties of a gas
electron multiplier (GEM) foil in pure dimethyl ether (DME) at low
pressures. The GEM is made from copper-clad liquid crystal polymer
insulator (LCP-GEM) designed for space use, and is applied to a time
projection chamber filled with low-pressure DME gas to observe the linear
polarization of cosmic X-rays. We have measured gains of a 100~\um-thick
LCP-GEM as a function of the voltage between GEM electrodes at various
gas pressures ranging from 10 to 190~Torr with 6.4 keV X-rays.
The highest gain at 190~Torr is about $2 \times 10^{4}$, while that
at 20~Torr is about 500. We find that the pressure and electric-field
dependence of the GEM gain is described by the first Townsend coefficient.
The energy scale from 4.5 to 8.0 keV is linear with non-linearity of
less than 1.4$\%$ above 30~Torr.

}

\begin{document} 
\section{Introduction}\label{sec:Intro}
A Gas Electron Multiplier (GEM) invented by F. Sauli in 1996 \cite{bib1} is one of the micro-pattern gas detectors. 
We have developed GEMs which are designed for space use and consist of copper electrodes and a liquid crystal polymer insulator (LCP-GEM) using laser etching technique \cite{bib2,bib3}. 
The effective gain of the LCP-GEM with a hole pitch of 140~\um, a hole diameter of 70~\um, and a thickness of 100~\um~reaches $10^4$ at an applied voltage of 720~V in mixture of 70\%~Ar and 30\%~CO$_2$ at 1~atm \cite{bib3}. 
The LCP-GEM shows a stable gain with time variation of about 0.5\% for 3 hours since the high voltage is applied to the LCP-GEM electrodes \cite{bib3}.
In addition, the LCP-GEM shows a good gain uniformity across the whole active surface with a standard deviation of about 4\%~\cite{bib3,Asami}. 
The distribution of hole diameters across the LCP-GEM is homogeneous with a standard deviation of about 3\%~\cite{bib3,Asami}.

The primary purpose of our LCP-GEM development is to construct a photoelectric X-ray polarimeter on-board the NASA's sounding rocket, XACT, the first dedicated mission for high-sensitive observation of cosmic X-ray polarization \cite{XACT}.
To detect linear X-ray polarization with the photoelectric effect which is the dominant interaction in the X-ray region, it is necessary to know the direction of photoelectrons emission.
As a result of the photoionization of an atomic s-orbital electron, a photoelectron is ejected preferentially in the direction of the electric field of the incident X-ray \cite{bib4}. 
The polarimeter uses Time Projection Chamber (TPC) technique with strip readout to image the photoelectron tracks. 
Drifting electrons forming the photoelectron track are multiplied by the single LCP-GEM and then are collected by strip electrodes with 121~\um~pitch and read by APV25 \cite{APV25} with 20~MHz sampling.
The GEM gain required for the polarimeter is 3000.
The polarimeter also has the spectroscopic ability with an energy resolution of $<20$\% in FWHM at 6.4 keV by reading signals from the GEM cathode.
The detailed design of the polarimeter is described elsewhere \cite{bib4}. 

Pure dimethyl ether (DME) is selected as a detector gas because of its slow drift velocity and small diffusion of drift electrons \cite{bib5}. 
The high-sensitive measurement of X-ray polarization requires fine photoelectron track images.
Therefore, a low-pressure gas is suitable to extend the track length.
On the other hand, a low-pressure gas decreases the X-ray detection efficiency of the polarimeter.
We anticipate that the optimum gas pressure for DME ranges from 50 to 150~Torr in consideration of trade-off between detected count rate and modulation factor~\cite{XACT}. 
However LCP-GEMs have never been operated below 190~Torr.
In those low gas pressures, discharge is one of the most significant risk to operate a GEM.

In this paper, we report systematic studies of LCP-GEM gains in pure DME below 190~Torr under stable performance of electronic amplification  without discharge.
Section \ref{sec:Setup} details the experimental setup and procedure.
We describe gain curves at various low pressures in Section \ref{sec:gain}, explanation of the gain behavior with the first Townsend coefficient in Section \ref{sec:alpha} and energy scale linearity in Section \ref{sec:linearity}.
We show normal operation limit without discharge for LCP-GEM at lower pressure in Section \ref{sec:10TorrSig}.
Lastly, we summarize the observation results in Section \ref{sec:summary}.

%
%
\section{Experimental setup and procedure}\label{sec:Setup}
\begin{figure}[htbp]
  \begin{tabular}{cc}
    \begin{minipage}{0.5\hsize}
      \begin{center}
    \includegraphics[width=8.0cm]{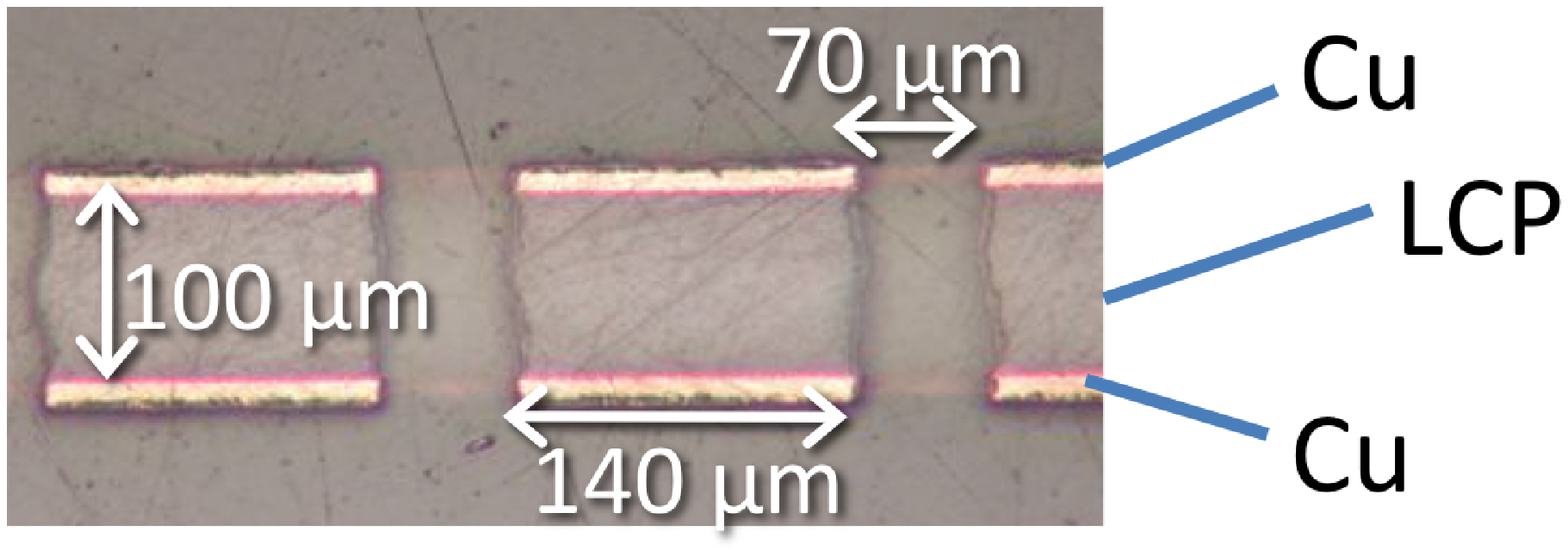}
     \caption{Cross-section of the LCP-GEM foil.}
      \label{fig:setup1}
      \end{center}
    \end{minipage}
\hspace{-0.5mm}
\begin{minipage}{0.5\hsize}
\begin{center}
    \includegraphics[width=6cm]{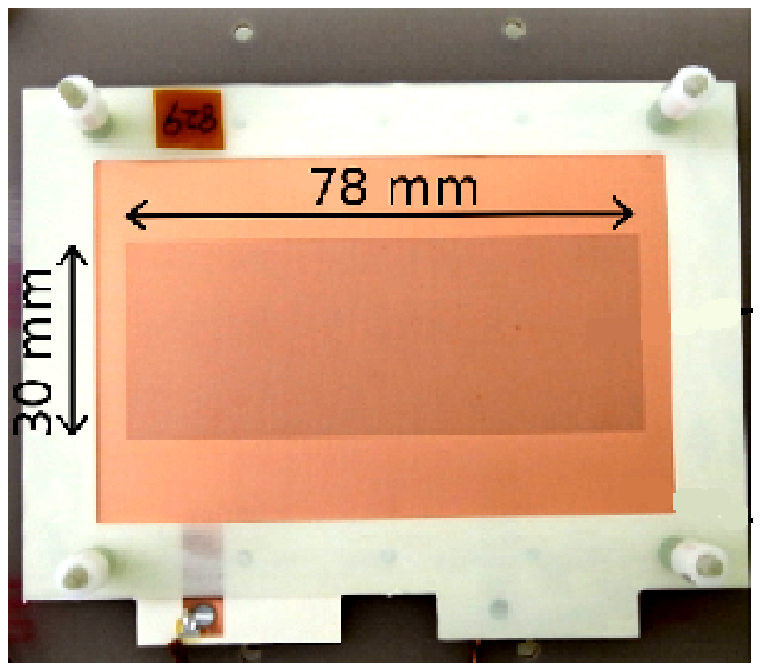}
    \caption{Top-view of the LCP-GEM foil.}
    \label{fig:setup2}
\end{center}
\end{minipage}
 \end{tabular}
\end{figure}

The LCP-GEM foil we used in this work was the identical geometry to the flight one for the XACT rocket mission with the hole pitch of 140~\um, the hole diameter of 70~\um, the thickness of 100~\um, and the active area of $78\times30~{\rm mm}^2$.
The cross-section and top view of the LCP-GEM are shown in Figures \ref{fig:setup1} and \ref{fig:setup2}, respectively.

\begin{figure}[tbp] 
\centering
    \includegraphics[width=15.0cm]{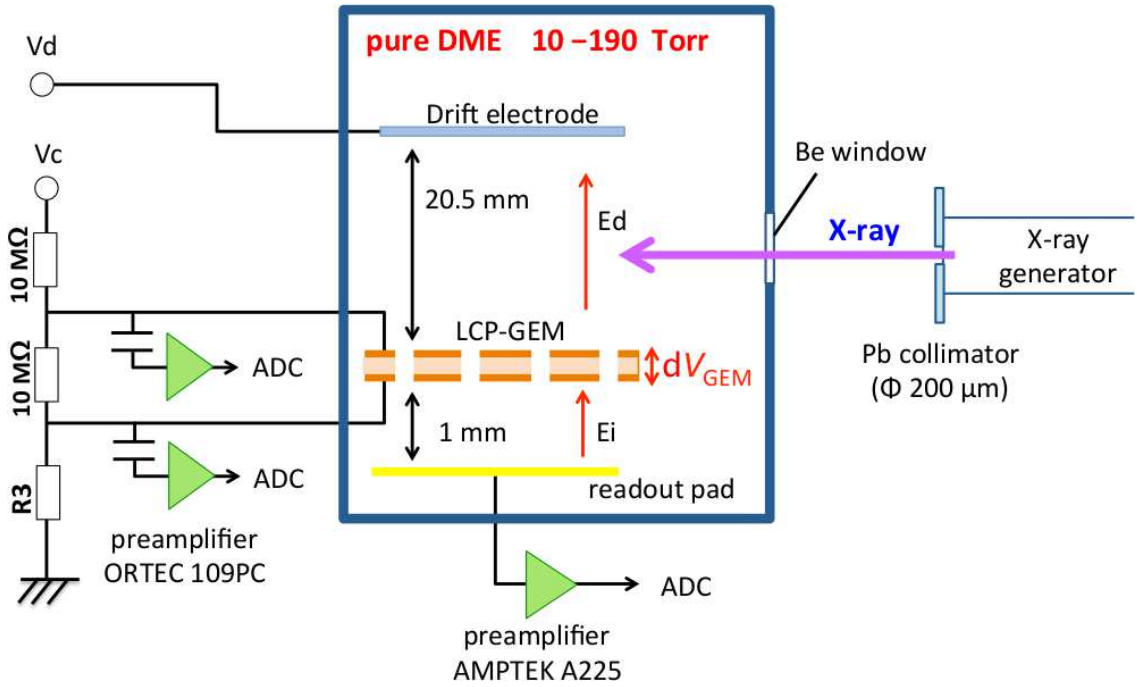}
\caption{Schematic view of the LCP-GEM experimental setup.}
\label{fig:setup3}
\end{figure}

Figure \ref{fig:setup3} shows a schematic view of the LCP-GEM test setup.
The chamber was filled with DME gas with purity of 99.99$\%$ and consisted of a drift plane, the LCP-GEM foil, and a readout pad, all of which had the active area of $78 \times 30~{\rm mm}^2$.
The gap between the drift plane and LCP-GEM was 20.5~mm and that between the LCP-GEM and readout pad was 1.0~mm.
A high voltage was applied to the drift plane with a power supply, REPIC RPH-042. 
The electric field in the drift region (\Ed) at each gas pressure ($P$) was adjusted based on the numerically predicted relation with Magboltz \cite{magboltz},  \Ed$=196 \times P/(190~{\rm Torr})~{\rm V~cm}^{-1}$, so that the electron drift velocity becomes a constant value of 0.24~\um~ns$^{-1}$, which is the same velocity as that for the rocket mission. 
The high voltages applied to the GEM electrodes were supplied by another REPIC RPH-042 with a voltage divider consisting of three resistors, one of which (R3 in Figure \ref{fig:setup3}) was adjustable to avoid electron amplification in the induction region by limiting the electric field in that region (\Ei) below \Ei$<4000{\times}P/(190~{\rm Torr})~{\rm V~cm}^{-1}$ \cite{take}.
X-rays from an X-ray generator were collimated to a diameter of 200~\um~with a lead collimator.
The beam was introduced into the chamber through the a Beryllium window in parallel to the LCP-GEM at the drift height from the GEM cathode of 1~mm.

The detailed characteristics of the preamplifiers and main amplifiers for the readout pad and GEM electrodes are summarized in Table \ref{tab:amp}.
The custom-made amplifier for readout pad signals had a discriminator and gate generator to activate a trigger for a data acquisition system.
Individual amplified pulse heights of the readout pad, GEM cathode, and GEM anode were simultaneously converted to digital value by a VME peak-hold ADC (Clear Pulse 1113A) in synchronization with the trigger, and were recorded in a hard disk with a Linux machine. 

Before the gain measurement, we determined conversion factor of individual signal readout systems from the ADC channel to the charge amount in coulomb by inputting test pulses to the preamplifiers.
Next, we measured GEM gain curves at various gas pressure ranging from 10 to 190~Torr with 6.4~keV (Fe-K$\alpha$) X-rays.
Then we evaluated energy scale linearity at different gas pressure with 4.5 (Ti-K$\alpha$), 6.4 and 8.0~keV (Cu-K$\alpha$) by changing a target material in the X-ray generator.
We collected 30000 X-ray events at each measurement.
At each gas pressure, we measured the GEM gain with gradual increase of the GEM voltage and stopped the gain measurement to prevent unexpected breakdown due to the rapid increase of micro-discharges when the total count of overflow events caused by discharge exceeded 30 counts, 0.1\% of total count of the collected events. 

\begin{table}[htbp]
\caption{Characteristics of the preamplifiers and main amplifiers.} 
\begin{center}
\begin{tabular}{cccl}
\hline
			& Type		 & Name				&Specification\\
\hline
\hline
			    &       			&             		& charge-sensitive amplifier\\
	 		    &				& 				& no shaper \\
			    &preamplifier	& ORTEC 109PC	&rise time : 20 ns @ 0~pF \\
GEM anode	    &       			&             		&decay time : 50 \us  \\
GEM cathode	    &       			&             		&test input capacitor : 1~pF \\ \cline{2-4} 			    		    				
	 		    &			 	&				& shaping amplifier \\
			    &main amplifier& ORTEC 572  	&shaping time : 6 \us \\
			    &       			&             		&gain :  20\\ 			    		    				
\hline
			    &       			&             		& charge-sensitive amplifier\\
	 		    &	preamplifier 	&AMPTEK A225	& shaping amplifier \\
			    &				& 				&peaking time : 2.4 \us \\ 
readout pad	    & 			& 		           	&test input capacitor : 2~pF\\ \cline{2-4} 	
	 		    &			 	&				& no shaper \\			    
			    &main amplifier &custom-made 	&with a discriminator and gate generator \\		
    	 		    &			 	&				&gain : 20 or 1 \\			
\hline
\end{tabular}
\end{center}
\label{tab:amp}
\end{table}

%
%
\section{Results $\&$ Discussion}
\label{sec:results}
\subsection{Gain curves at different low pressures}
\label{sec:gain}
\begin{figure}[tbp] 
\centering
\includegraphics[width=8.0cm, trim = 0 0 0 21, clip]{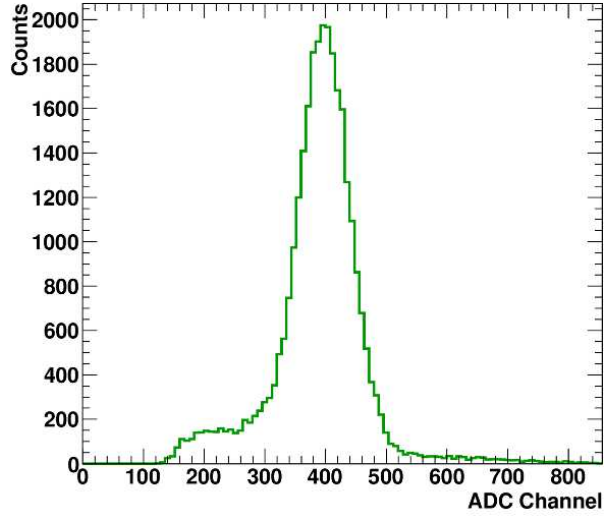}
\caption{ADC spectrum of the signals read from the readout pad at \dV$=400$~V and in 190~Torr of pure DME irradiated with 6.4~keV X-rays. The energy resolution was 20\% (FWHM).}
\label{fig:spec}
\end{figure}

Figure \ref{fig:spec} shows an ADC spectrum obtained from the readout pad when the chamber is irradiated with 6.4~keV X-rays at gas pressure of 190~Torr and the GEM voltage (\dV) of 400~V.
The spectral peak has the low-energy tail which is thought to be due to energy loss  caused by photoelectron collisions with the GEM.
We fitted a Gaussian model to the spectrum and got the energy resolution of 20\% in FWHM.

We derived the effective GEM gain ($G_{\rm eff}$) from the following formula:
\begin{eqnarray}
G_{\rm eff} = \frac{Q_{\rm mean}}{e n_e},
\end{eqnarray}
where $Q_{\rm mean}$ is the total amount of charges induced in the readout pad calculated from the best-fit peak channel of the Gaussian model, $e$ is the electron charge of $1.602\times10^{-19}$~C,  and $n_e$ is the total number of electron-ion pairs created by an incident X-ray.
We employed $n_e=268$ for 6.4~keV X-rays in pure DME~\cite{bib5}.

\begin{figure}[htbp]
  \begin{tabular}{cc}
    \begin{minipage}{0.5\hsize}
      \begin{center}
	\includegraphics[width=8.2cm, trim = 38 5 40 0, clip]{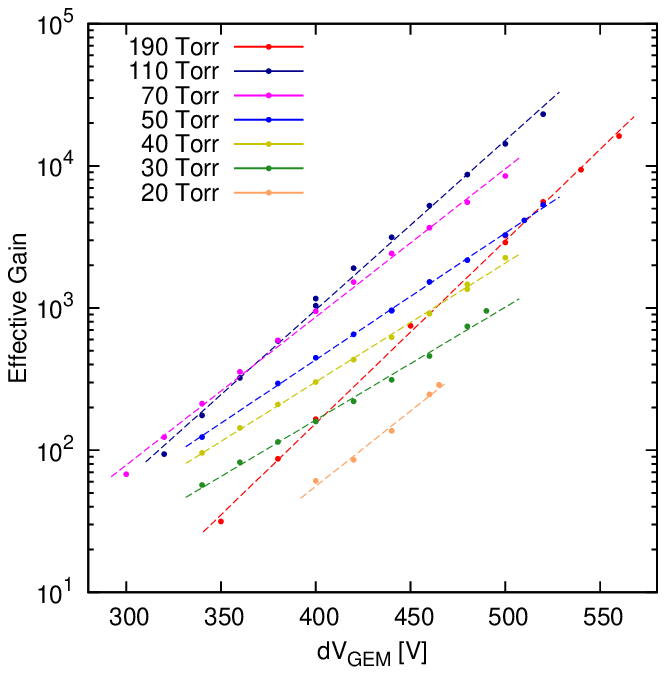}
	\caption{Effective gain of LCP-GEM in pure DME as  a function of GEM voltages at different gas pressures below 190~Torr with 1$\sigma$ error bar. Errors of measured charge are less than 0.6$\%$. The straight lines are plotted to represent an exponential function for convenience.}
	\label{fig:gain}
      \end{center}
    \end{minipage}
\hspace{1mm}
\begin{minipage}{0.5\hsize}
\begin{center}
\includegraphics[width=8.2cm, trim = 38 5 40 0, clip]{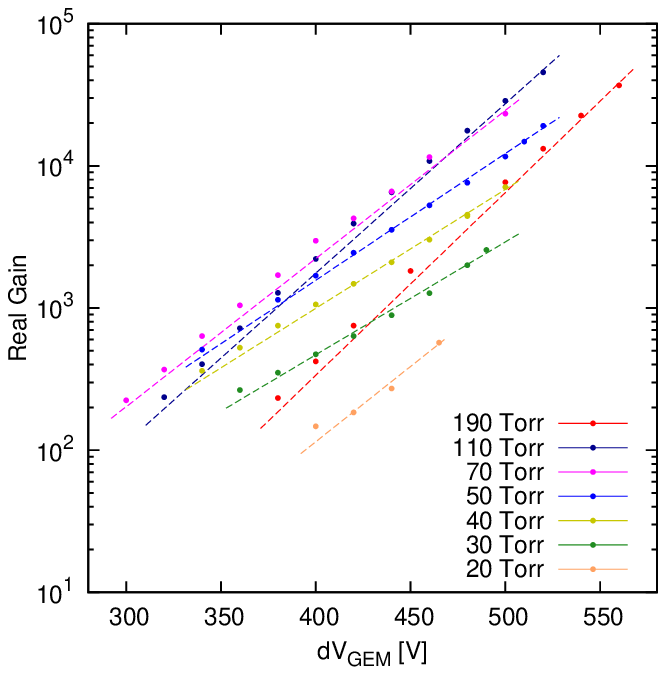}
\caption{Same as Figure \protect\ref{fig:gain}, but the y-axis is the real gain of LCP-GEM.
Errors of real gain are less than 0.81$\%$.
The straight lines are plotted to represent an exponential function for convenience. The slope value of each straight line is the same as that in Figure~\protect\ref{fig:gain}.}
\label{fig:real_gain}
      \end{center}
    \end{minipage}
 \end{tabular}
    \end{figure}

Figure \ref{fig:gain} shows effective gain curves of the LCP-GEM at different gas pressures from 20 to 190~Torr.
The highest gain under stable operation at 190~Torr is $2\times10^4$ at \dV$=560$~V, while that at 20~Torr is $\sim$ 300 at \dV$=470$~V.
The highest gain above 50~Torr exceeds 3000, which meets the polarimeter requirement.
Although the gain curve at 190~Torr can be reproduced by an exponential function. the others at lower pressure show deviation from it.

In addition, we determined the real GEM gain ($G_{\rm real}$) derived from the sum of charge amounts induced in the readout pad and GEM anode using the same data set.
The real gain represents the amplification degree for electrons drifting in the GEM hole, while the effective gain is the real gain multiplied by the amplified electron collection efficiency of the readout pad \cite{Benlloch}.
Figure \ref{fig:real_gain} shows real gain curves at different gas pressures from 20 to 190~Torr.
The real gain is approximately twice of the effective one because the charge amount of the GEM anode is almost the same as that of the readout pad.
The highest gain at 190~Torr is $4\times10^4$, while that at 20~Torr is $\sim$ 600.
These real gain curves also show deviation from an exponential function.

\subsection{Comparison of the first Townsend coefficient}
\label{sec:alpha}
To comprehensively characterize the gain variations with different gas pressures and GEM voltages,
we derived the first Townsend coefficient, $\alpha$, given by 
\begin{eqnarray}
\alpha = \frac{{\rm ln}(G_{\rm real})}{x}, 
\end{eqnarray}
where $x$ is the amplification length for electrons.
Here, $x$ was assumed to be the GEM thickness of 100~\um.

\begin{figure}[tbph] 
\centering
\includegraphics[width=13.0cm]{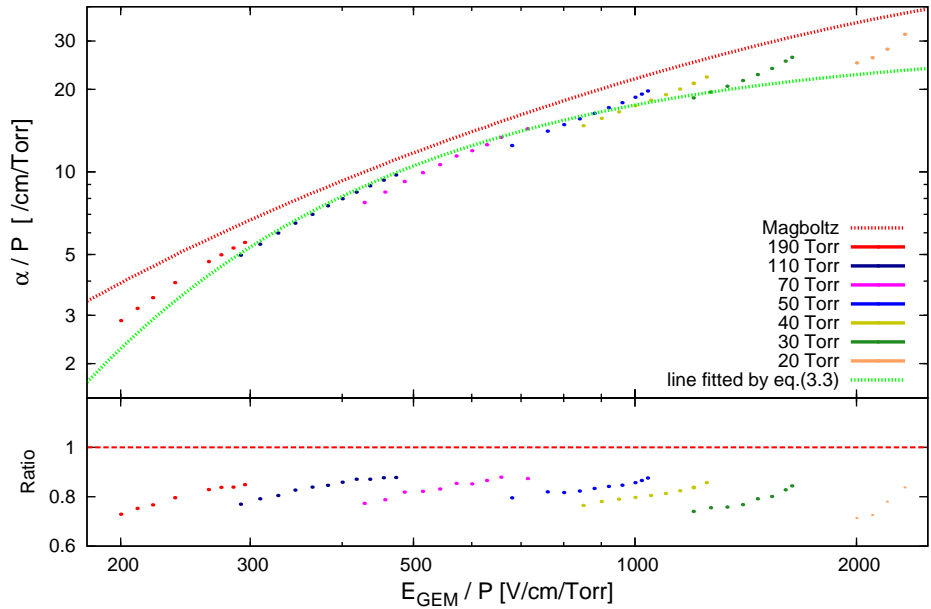}
\caption{Top: Comparison between measured $\alpha$ and $\alpha$ simulated by Magboltz \cite{magboltz}.
Bottom: Ratio between Magboltz and this experiment with 1$\sigma$ error bar. The pressure range is 10--760~Torr, and range of the electric field is 10--60~kV/cm in Magboltz simulation.}
\label{fig:Tow_MAG}
\end{figure}

Figure \ref{fig:Tow_MAG} shows the observed $\alpha$ as a function of $E_{\rm GEM}/P$ superposed with Magboltz \cite{magboltz} predictions. 
$E_{\rm GEM}$ is the electric field applied to the GEM and was estimated from the GEM operating voltage divided by the GEM thickness of 100~\um.
The Magboltz simulation assumes a uniform electric field.
The data points is roughly reproduced with an exponential function, although they show deviation from it in the higher $E_{\rm GEM}/P$ range. 
This is because DME ions gain  a high enough kinetic energy from such a strong electric field to ionize DME molecules and emit additional electrons which are amplified in the GEM hole.
The measured $\alpha$ values were $\sim$ 80$\%$ of Magboltz ones, suggesting that the distance of electron amplification in the GEM hole was not the GEM thickness of 100~\um~but $\sim$ 80~\um.
The suggestion is supported by Ref~\cite{sim} who showed that the avalanche distance for electron is shorter than the GEM thickness by a Garfield simulation.
The residuals from the exponential model change with the electric field in the bottom panel of Figure~\ref{fig:Tow_MAG}.
One reason is that the distance of electron amplification in the GEM hole depends on the GEM operating voltage.
Another reason is that the measured charge depends on the induction electric field (\Ei) which changed at each gain measurement by adjusting the resistor R3 in Figure~\ref{fig:setup3}.

Sharma et al. (1993) \cite{alpha} studied the behavior of $\alpha/P$ vs. $E/P$ in pure DME and fitted the following empirical formula:
\begin{eqnarray}
\frac{\alpha}{P} = A~{\rm exp}\Bigl( -B~\frac{P}{E}\Bigr),
\end{eqnarray}
where $A$ and $B$ are free parameters.
Sharma et al. (1993) reported $A=8.0$ and $B=213.1$ for their measurements.
We note that the measurements of \cite{alpha} were carried out with a parallel plate in $E/P$ range of 20--30 V~cm$^{-1}$~Torr$^{-1}$, which was $\sim10$ times lower than our measurements.
With the same way descried in Ref~\cite{alpha}, we fitted the empirical formula to the data points.
The best-fit values of $A$ and $B$ are respectively  $29.22\pm0.52$ and $511\pm10$, both of which are a few times larger than the values in Sharma.
A possible explanation of these differences is that the electric field in the GEM holed is more complex than that of the parallel plate.

\subsection{Linearity of Energy Scale}\label{sec:linearity}
\begin{figure}[tbph] 
\centering
\includegraphics[width=13.0cm, trim = 0 0 0 15, clip]{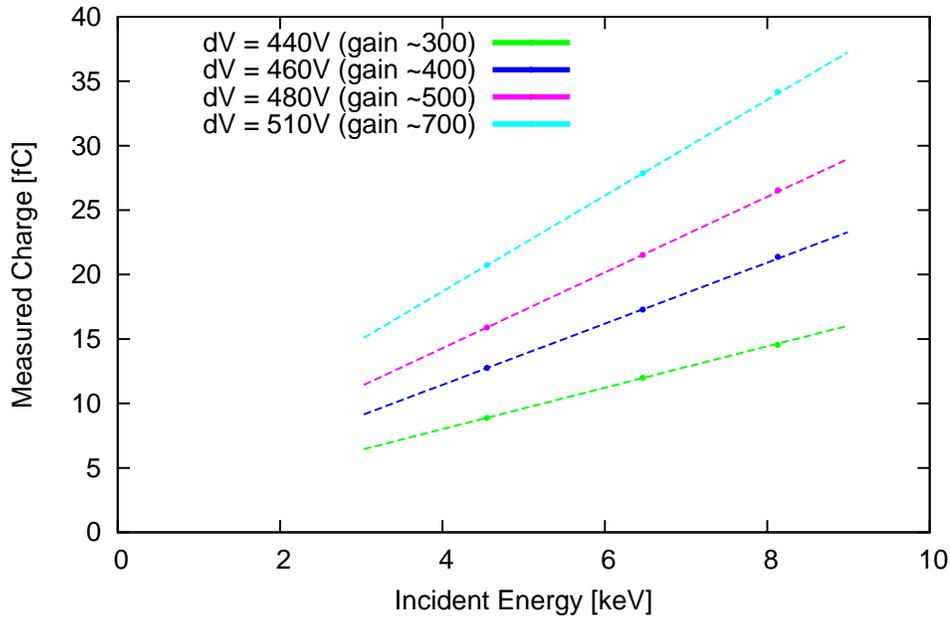}
\caption{Measured charge as a function of incident X-ray energy at 30~Torr in pure DME with 1$\sigma$ error bar. These data were fitted by straight line. Errors of measured charge are less than 0.2$\%$.}
\label{fig:EnergyScale}
\end{figure}

In order to verify that the LCP-GEM operates in the proportional region below 190~Torr, we measured the linearity of energy scale with 4.5, 6.4, and 8.0~keV X-rays.
Figure \ref{fig:EnergyScale} shows the charge amount of the readout pad at 30~Torr.
The non-linearity of the energy scale is less than 1.4$\%$, indicating that the LCP-GEM operated in the proportional region at 30~Torr.
In addition, the linearity at 50, 70, 110 and 190~Torr was confirmed with the same way as that at 30~Torr.
 
\subsection{Signal shapes at 10 Torr}\label{sec:10TorrSig}
\begin{figure}[htbp]
  \begin{tabular}{cc}
    \begin{minipage}{0.5\hsize}
      \begin{center}
        \includegraphics[width=7.0cm]{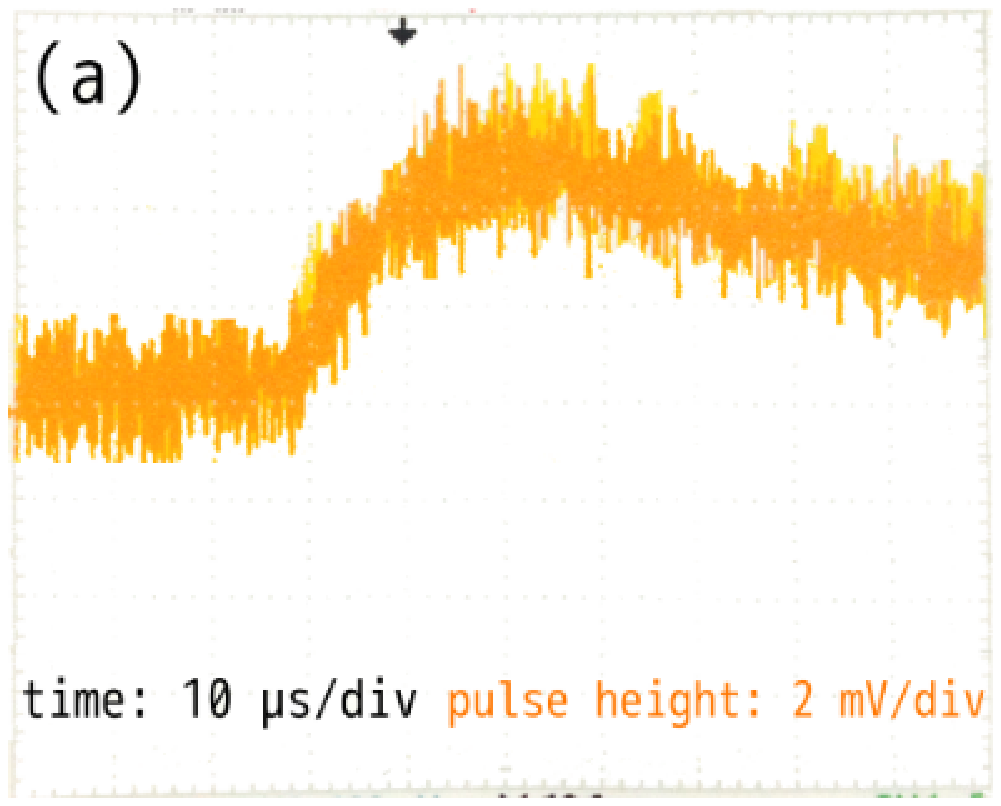}
      \end{center}
     \end{minipage}
     \begin{minipage}{0.5\hsize}
      \begin{center}
	  \includegraphics[width=7.0cm]{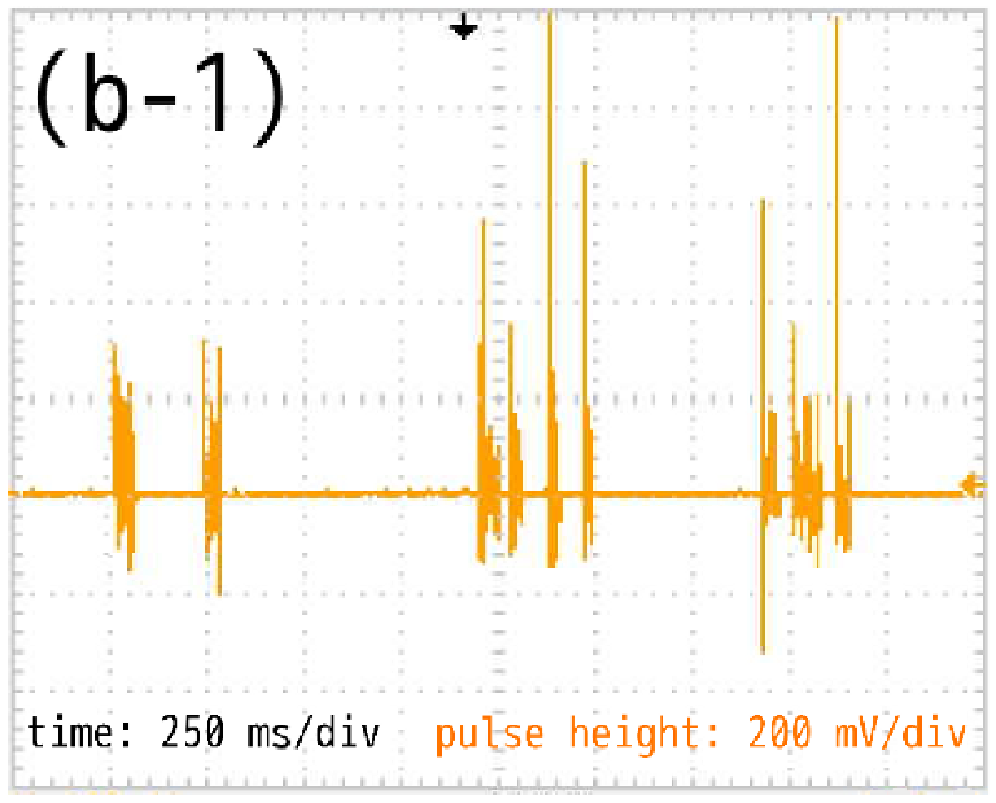}
      \end{center}
    \end{minipage}
\\
\\
\begin{minipage}{0.5\hsize}
\begin{center}
  \includegraphics[width=7.0cm]{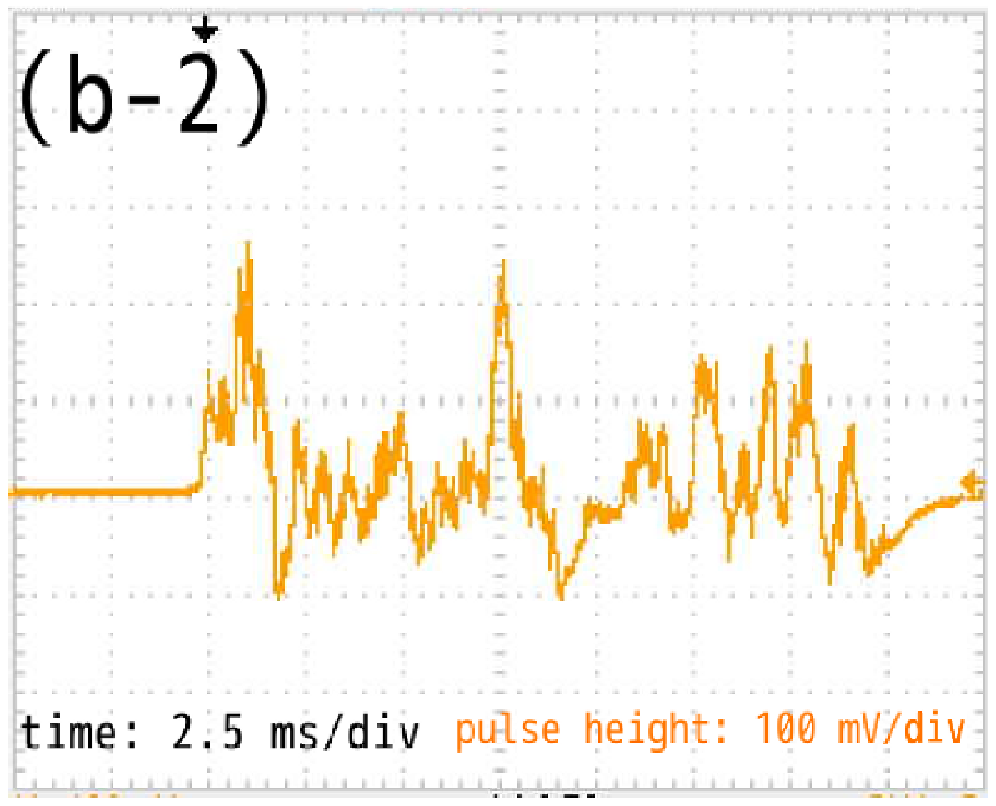}
  \end{center}
\end{minipage}
\begin{minipage}{0.5\hsize}
  \begin{center}
  \includegraphics[width=7.0cm]{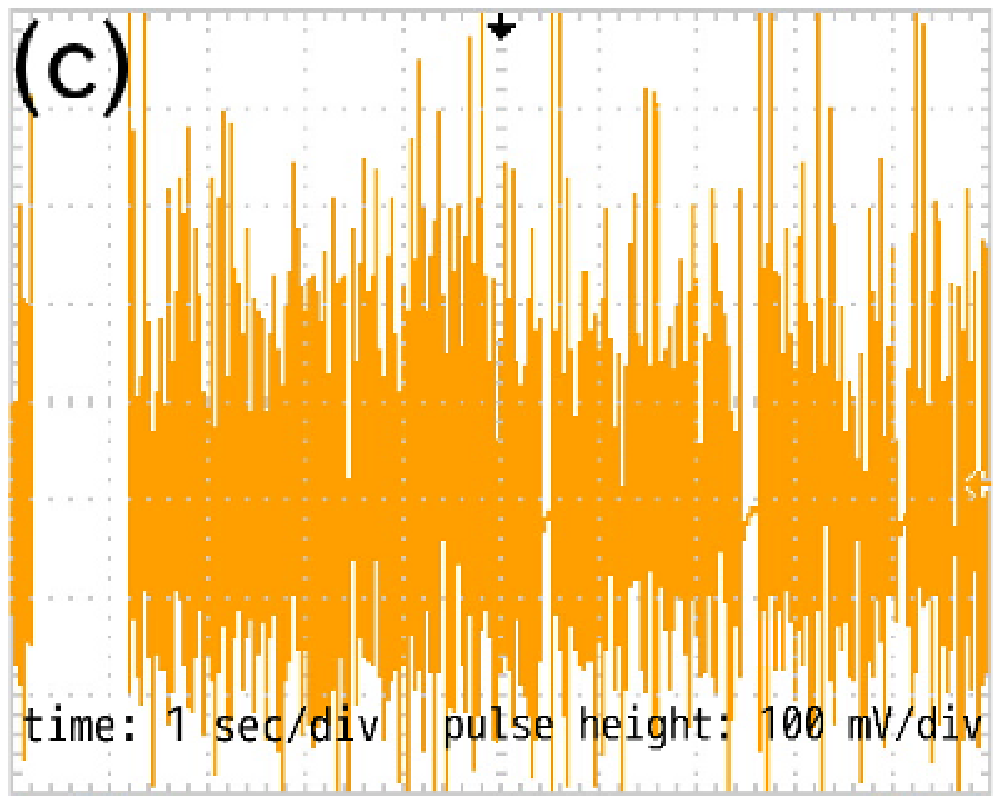}
  \end{center}
\end{minipage}
\end{tabular}
\caption{The preamplifier output signal from the readout pad with \Ed~of 10.3~V/cm in pure DME at 10~Torr. (a)\dV$=430$~V, (b-1)\dV$=437$~V, (b-2) closeup of time scale and (c)\dV$=440$~V. The preamplifier used for these measurements was not AMPTEK A225, but ORTEC 109PC to check signal without shaping. }
\label{fig:PulseProfile}
\end{figure}

We investigated the LCP-GEM performance at lower pressure, 10~Torr.
Figure \ref{fig:PulseProfile} shows waveform of the ORTEC 109PC preamplifier output from the readout pad. 
Because the AMPTEK A225, which was used for the gain and energy scale measurements, contained the shaping amplifier,
it was replaced with ORTEC 109PC without shaping function to check signal waveform as raw as possible.
At \dV~of 430~V, the signal waveform with the duration about 10~\us~and the pulse height of about 4~mV was similar to those at 30~Torr.
However, the signal shape dramatically changed above 437~V.
At \dV~of 437~V, the signal had a much longer duration of about 10~ms, and a much lager pulse height of 200~mV, compared to the normal signal at 430~V.
Furthermore, the signal duration at \dV$=440$~V got longer than that at 437~V, while the pulse height did not change.
These signals disappeared when X-ray irradiation stopped, showing that they were induced by incident X-rays from the X-ray generator.

According to the gas ionization curve, the operation mode of a gas chamber transits from the proportional region to the Geiger region as the voltage goes up and/or the pressure goes down \cite{Knoll}.
The signal duration in the Geiger region takes just a few 100~\us~in a Geiger counter because the ultraviolet photons produced in the original avalanche cause consecutive avalanches in the region \cite{Knoll}.
The feature of the long signal in the Geiger region is similar to the measured one at  \dV~$>437$~V and 10~Torr. 
Therefore, it appears likely that the operation mode of LCP-GEM transits from the proportional to the Geiger region at \dV~$>437$~V and 10~Torr.
Further studies are needed to confirm our understanding.

%
%
\section{Summary }\label{sec:summary}
We have measured the gain properties of the LCP-GEM in pure DME gas at various low pressures below 190 Torr.
Here is the summary of the experimental results:

\begin{list}{$\cdot$}{}
\item[$\bullet$] We measured the LCP-GEM gain curves at 20--190~Torr. The effective gain at 190~Torr reached $\sim 2\times10^4$ at \dV$ = 560$~V without any micro-discharge, while that at 20~Torr was $\sim300$ at \dV$=470$~V.

\item[$\bullet$] The LCP-GEM gains at various gas pressures and GEM voltages were characterized by the first Townsend coefficient $\alpha$.

\item[$\bullet$]We found that measured $\alpha/P$ values were consistent with those simulated with Magboltz assuming that amplification length for electrons was $\sim$80~\um~which was 80\% of the actual GEM thickness. 

\item[$\bullet$] The energy scale from 4.5 to 8.0~keV was linear in the pressure range of 30 to 190~Torr.

\item[$\bullet$] The signal waveform at 10~Torr got long ($\sim10$~ms) and high ($\sim200$~mV) above the GEM voltage of 437~V, suggesting transition from proportional to Geiger mode of the LCP-GEM.

\end{list}

%
%
\acknowledgments
This work was supported by JSPS KAKENHI Grant Number 22244034.
Y.T. was supported by Grant-in-Aid for he Promotion of Science (JSPS) Fellows (No.25$\cdot$5448).
T.K. was supported by Japan Society for JSPS Grant-in-Aid for Young Scientists (B) (No.24740185).


\begin{thebibliography}{9}

\bibitem{bib1}
F. Sauli, 
\emph{GEM: A new concept for electron amplification in gas detectors}
{\emph{Nucl. Instr. and Meth.} {\bf A386} (1997) 531}.


\bibitem{bib2}
T. Tamagawa, et al.,
\emph{Fine-pitch and thick-foil gas electron multipliers for cosmic x-ray polarimeters},
{\emph{Nucl. Instr. and Meth.} {\bf  A560} (2006) 418}.

\bibitem{bib3}
T. Tamagawa, et al.,
\emph{Development of thick-foil and fine-pitch GEMs with a laser etching technique},
{\emph{Nucl. Instr. and Meth.} {\bf  A608} (2009) 390}.


\bibitem{Asami}
F. Asami, et al.,
\emph{Mapping study of gain and hole diameter of Japanese GEMs},
{\emph{Nuclear Science Symposium Conference Record (NSS/MIC), 2009 IEEE}}.


\bibitem{XACT}
K. Gendreau, et al.,
\emph{The x-ray advanced concepts testbed (XACT) sounding rocket payload},
\emph{Proc. of SPIE Vol. 8443 84434V-1(2012)}.


\bibitem{bib4}
J.K. Black, et al., 
\emph{X-ray polarimetry with a micropattern TPC}
{\emph{Nucl. Instr. and Meth.} {\bf A581} (2007) 755}.


\bibitem{APV25}
M.J. French et al.,
\emph{Design and results from the APV25, a deep sub-micron CMOS front-end chip for the CMS tracker}
{\emph{Nucl. Instr. and Meth.} {\bf A466} (2001) 359}.


\bibitem{bib5}
A. Sharma,
\emph{Properties of some gas mixtures used in tracking detectors},
{\emph{ICFA} {\bf 16} (1998)}.


\bibitem{magboltz}
S.F. Biagi,
\emph{A description of the Magboltz program, with results compared to experiment.},
{\emph{Nucl. Instr. and Meth.} {\bf A421} (1999) 234-240}.



\bibitem{take}
Y. Takeuchi et al.,
\emph{Signal shape and charge sharing between electrodes of GEM in dimethyl ether}
{\emph{JINST} {\bf 7 C03042} (2012) 2ND INTERNATIONAL CONFERENCE ON MICRO PATTERN GASEOUS DETECTORS (MPGD2011) proceedings}.




\bibitem{Benlloch}
J. Benlloch et al., 
\emph{Further developments of the gas electron multiplier (GEM)},
{\emph{Nucl. Instrum. Meth.} {\bf A 419} (1998) 410}


\bibitem{sim}
O. Bouianov, et al.,
\emph{Foil geometry e!ects on GEM characteristics},
{\emph{Nucl. Instr. and Meth.} {\bf A458} (2001) 698}.


\bibitem{alpha}
A. Sharma and F. Sauli,
\emph{First Townsend coefficient measured in argon based mixtures at high fields},
{\emph{Nucl. Instr. and Meth.} {\bf A334} (1993) 420}.


\bibitem{Knoll}
G.F. Knoll,
\emph{Radiation detection and measurements, 3rd edition},
John Wiley and Sons, Inc., New York 2000


\end{thebibliography}
\end{document}